\documentclass[preprint,aps,nofootinbib]{revtex4-1}
\pdfoutput=1

\usepackage{amsmath}  
\usepackage{amsbsy,amssymb,amsfonts}
\usepackage{graphicx}
\usepackage{color}
\usepackage{epsf}
\usepackage[utf8x]{inputenc}
\usepackage{hyperref}
\usepackage{multirow}
\def\bravert{\egroup\,\vrule\,\bgroup}

{\catcode`\|=\active
  \gdef\Twoint#1{\left(\mathcode`\|"8000\let|\bravert {#1}\right)}}
{\catcode`\|=\active
  \gdef\Braket#1{\left<\mathcode`\|"8000\let|\bravert {#1}\right>}}

\newcommand{\beq}{\begin{equation}}
\newcommand{\eeq}{\end{equation}}
\newcommand{\beqa}{\begin{eqnarray}}
\newcommand{\eeqa}{\end{eqnarray}}
\newcommand{\bea}{\begin{array}}
\newcommand{\eea}{\end{array}}

\newcommand{\bef}{\begin{figure}}
\newcommand{\ef}{\end{figure}}
\newcommand{\bc}{\begin{center}}
\newcommand{\ec}{\end{center}}
\newcommand{\bt}{\begin{table}}
\newcommand{\et}{\end{table}}
\newcommand{\btb}{\begin{tabular}}
\newcommand{\etb}{\end{tabular}}

\newcommand{\au}{{\em a.u.}}

\def\au{{\it a.u.\ }}
\begin{document}

\title { Atomic Static Electric Quadrupole Polarizability from Relativistic Many-Body Theory With Application To
         Thulium $\left(6s^2\, 4f^{13}\right)$\, $^2F_{7/2}$ and $^2F_{5/2}$ }

\vspace*{1cm}

\author{Timo Fleig}
\email{timo.fleig@irsamc.ups-tlse.fr}
\affiliation{Laboratoire de Chimie et Physique Quantiques,
             FeRMI, Universit{\'e} Paul Sabatier Toulouse III,
             118 Route de Narbonne, 
             F-31062 Toulouse, France }
\vspace*{1cm}
\date{\today}

\vspace*{1cm}
\begin{abstract}
In the present work a method for calculating atomic static electric rank-$n$ tensor quadrupole polarizabilities 
$\alpha_Q^{(n)}(J)$ for arbitrary atomic electronic structure is presented. It is based on a linear-symmetry 
representation of atomic states using large-scale relativistic configuration interaction wavefunctions, and the
quadrupole polarizabilities are obtained by a sum-over-states approach. 

The major result from applying this method
to atomic thulium is $\alpha_Q^{(0)}(\frac{7}{2}) = 2496^{+0}_{-231}$ \au for the electronic ground term
$^2F_{7/2}$ static electric scalar quadrupole polarizability which is roughly a factor of four smaller than same
property in the electronic ground state of the cesium atom. This finding adds to the favorable properties of
thulium atoms in atomic-clock research.
\end{abstract}

\maketitle
\section{Introduction}
\label{SEC:INTRO}
Stray electric-field gradients (EFG) are one of the components limiting the accuracy of atomic clocks
\cite{ludlow_rmp_2015,Itano_2000}.
The predominant resulting effects are through interaction with the atomic electric quadrupole moment
\cite{Oskay_QuaSupp2005} and by polarizing the atom through its electric quadrupole (E2) polarizability
\cite{Higgins_2021}.
The particular limitations to the accuracy of high-precision measurements and optical atomic clocks
and possible remedies have been studied and discussed recently for relevant atomic states supporting 
an electric quadrupole moment \cite{Arnold_Barrett_2019}.

The development of a transportable optical atomic clock using hyperfine levels in electronic ground-term
atomic thulium (Tm) for the clock transition has been furthered in recent years 
\cite{NatComm_Tm_2019,PRA_Tm_2016,PRA_Tm_2020,Atoms_Tm_2023,PRA_Tm_2024,Tm_JETPL_2025,Tm_PRXQ_2025,Tm_JETPL_2026}.
Among the contributions to an estimation of its uncertainty budget \cite{AIP_Tm_2020} are the electric
quadrupole moment \cite{Fleig_PhysRevA.107.032816}, the hyperfine Stark effect \cite{Fleig_PRA_2025},
and other contributions \cite{fritzsche_Tm_2024}
in the electronic ground-term components $^2F_{7/2}$ and $^2F_{5/2}$. The E2 polarizability of thulium,
however, remains unknown, and its calculation would be helpful in advancing the development of a high-precision
thulium atomic clock \cite{golovizin_private}.

The paper is organized as follows. In section \ref{SEC:THEORY} the theory of electric quadrupole polarizability
is briefly reviewed and a general method for its calculation in atomic states of arbitrary structure is
presented. This method is then applied in section \ref{SEC:RESULTS} first to the lithium atom where
highly accurate benchmark calculations are available and then to the thulium atom where genuine
predictions are made. The final section \ref{SEC:CONCL} is dedicated to concluding remarks.


\section{Theory}
\label{SEC:THEORY}
\subsection{Fundamental background}

Electric field-gradient quadrupole polarizability has been introduced for atomic and molecular systems by
Dalgarno and Buckingham \cite{dalgarno_1962,buckingham_1967}.
The second-order classical potential-energy shift of a charge distribution $\rho$ due to the presence of a 
static external electric field ${\cal E}$ is \cite{jackson}
\begin{equation}
 \Delta E_{\text{class.}} = - \frac{1}{6} \sum_{i,j}^3\, Q_{ij}
               \frac{\partial {\cal E}_{j,\text{ext}}({\bf{x}})}{\partial x_i}
 \label{EQ:DELTAE}
\end{equation}
where ${\bf{Q}}$ is the rank-2 cartesian tensor of electric quadrupole moment with components 
$Q_{ij} = \iiint\limits_{\cal{V}}\, \rho\, \left( 3x_i x_j\, - {\bf{x}}\,^2 \delta_{ij} \right) d{\cal{V}}$
and $\frac{\partial {\cal E}_{j,\text{ext}}({\bf{x}})}{\partial x_i}$ is an element of the rank-2 cartesian tensor of the 
electric-field gradient (EFG). Since ${\boldsymbol{\nabla}} \times {\cal E}_{\text{ext}} = {\bf{0}}$ in general, 
assuming ${\cal E}_{\text{ext}}$ 
to be axially symmetric reduces EFG to its diagonal and entails that 
$\frac{\partial {\cal E}_{1,\text{ext}}}{\partial x_1}=\frac{\partial {\cal E}_{2,\text{ext}}}{\partial x_2}$.
In addition, far from the charges producing ${\cal E}_{\text{ext}}$, $\boldsymbol{\nabla} \cdot {\cal E}_{\text{ext}} = 0$ 
is valid. From this then follows that $2\frac{\partial {\cal E}_{1,\text{ext}}}{\partial x_1} = 
\frac{\partial {\cal E}_{3,\text{ext}}}{\partial x_3}$. As a consequence, Eq. (\ref{EQ:DELTAE}) becomes
\begin{equation}
 \Delta E_{\text{class.}} = - \frac{1}{6} \left[ Q_{33} -\frac{1}{2} \left( Q_{11} + Q_{22} \right)
               \frac{\partial {\cal E}_{3,\text{ext}}({\bf{x}})}{\partial x_3} \right]
 \label{EQ:DELTAE1}
\end{equation}
The quadrupole moment tensor is manifestly traceless
\begin{equation}
 Q_{11} + Q_{22} + Q_{33} = 0
\end{equation}
from which Eq. (\ref{EQ:DELTAE1}) becomes
\begin{equation}
 \Delta E_{\text{class.}} = - \frac{1}{4} Q_{33} \frac{\partial {\cal E}_{3,\text{ext}}({\bf{x}})}{\partial x_3}
 \label{EQ:DELTAE2}
\end{equation}

\subsection{Static electric quadrupole polarizability}

For a quantum system the energy shift from Eq. (\ref{EQ:DELTAE2}) is represented by the corresponding Hamiltonian
\begin{equation}
 \hat{H}_Q = - \frac{1}{4} \hat{Q}_{33} \frac{\partial {\cal E}_{3,\text{ext}}({\bf{x}})}{\partial x_3}
\end{equation}
where the field is treated classically and the quadrupole-moment operator $\hat{Q}$ is introduced.
The ensuing second-order energy shift of the atomic state $n$ can then be written as
\begin{equation}
 \Delta E^{(2)}_n = \sum\limits_{k \ne n}^{N}\, 
    \frac{\left|\left< \psi_n \left| \hat{H}_Q \right| \psi_k \right>\right|^2}{E_n - E_k} 
  = \frac{1}{16} \left(\frac{\partial {\cal E}_{3,\text{ext}}({\bf{x}})}{\partial x_3} \right)^2
      \sum\limits_{k \ne n}^{N}\, 
    \frac{\left|\left< \psi_n \left| \hat{Q}_{33} \right| \psi_k \right>\right|^2}{E_n - E_k}
    \label{EQ:ENERGY_N_Q}
\end{equation}
where $N$ is the total number of atomic many-body states included in the summation.
Defining the static electric quadrupole polarizability in atomic state $n$ as
\begin{equation}
 \alpha_Q(n) = -2\sum\limits_{k \ne n}^{N}\,
    \frac{\left|\left< \psi_n \left| \hat{Q}_{33} \right| \psi_k \right>\right|^2}{E_n - E_k}
    \label{EQ:QPOLn}
\end{equation}
Eq. (\ref{EQ:ENERGY_N_Q}) becomes
\begin{equation}
 \Delta E^{(2)}_n = -\frac{1}{32} \alpha_Q(n)\, \left(\frac{\partial {\cal E}_{3,\text{ext}}({\bf{x}})}{\partial x_3} \right)^2
\end{equation}
If other conventions for the quadrupole-polarizability tensor (denoted $\alpha_{Q,n}$) are employed then the prefactor 
in the energy expression changes. For instance, with the conventions of Maroulis et al. it picks up a factor
of four and results in the equivalent often-cited \cite{Maroulis_1988} form
\begin{equation}
 \Delta E_n = -\frac{1}{8} \alpha_{Q,n} \left( {\bf\nabla} {\cal E} \right)^2
\end{equation}

\subsection{Present formalism}

In the present work the static field-gradient quadrupole polarizability from Eq. (\ref{EQ:QPOLn}) is calculated using
a sum over atomic states represented by wavefunctions transforming as to the irreducible representation $\Gamma$ of the
linear rotation double group with inversion center, $D^*_{\infty h}$. In practice, the subgroup $D^*_{32 h}$ is used
which for atomic states has large enough dimension to allow for completely isomorphic mappings of atomic states 
$\left| M_J \right>$ onto irreducible representations of $D^*_{32 h}$ \cite{fleig_gasci2}. 
This means that the valid quantum number for such states is $M_J$, the
projection quantum number of total electronic angular momentum $J$. Since the quadrupole-moment operator $\hat{Q}_{33}$
is totally symmetric there is a selection rule $\Delta M_J = M_J' - M_J = 0$ for matrix elements 
$\left< M_J | \hat{Q}_{33} | M_J' \right>$.

Eq. (\ref{EQ:QPOLn}) is consequently written as
\begin{equation}
 \label{EQ:ALPHAQ}
 \alpha_Q(\psi_0(M_J)) = -2\, \sum\limits_{k \ne 0}^{K-1}\, 
 \frac{\left|\left< \psi_0(M_J) \left| \hat{Q}_{33} \right| \psi_k(M_J) \right>\right|^2}{E_0 - E_k}
\end{equation}
where $\psi_0(M_J)$ is the target state,
$\hat{Q}_{33} = \hat{Q}_{zz} = -\frac{1}{2}\, \sum\limits_{i=1}^n\, \left( 2{\hat{z}}^2(i) - {\hat{x}}^2(i) 
- {\hat{y}}^2(i) \right)$ is a component of the rank-2 electric quadrupole moment tensor operator of
the $n$-electron system \cite{Sundholm_Olsen_QuaMom1993,Fleig_PhysRevA.107.032816}, 
$E_k$ is the unperturbed energy of many-body state $\psi_k$ 
and $K$ is the number of CI vectors of lowest total energy used in the summation over states, including
the target state.
Defining $N = dim\left(\Gamma(M_J)\right)$ with $\Gamma(M_J)$ the irreducible representation of the
target state for the quadrupole polarizability it is in practice always the case that $K << N$.
The matrix elements in the nominator of Eq. (\ref{EQ:ALPHAQ}) are calculated as discussed in Ref. 
\cite{Fleig_PhysRevA.107.032816}.
In the present work $\alpha_Q$ is represented in atomic units which are $a_0^5$.

The rank-$n$ tensor electric field-gradient quadrupole polarizabilities $\alpha^{(n)}_Q(J)$ for $J$ states can now 
be obtained with the help of Refs. \cite{Manakov_Ovsyannikov_1979,Manakov_Ovsyannikov_1986,Chen_Raithel_2015}.
There,
\begin{eqnarray}
 \nonumber
 \alpha_Q\left(\psi_0(M_J)\right) &=& \alpha_Q^{(0)}\left(J\right) + \frac{3M_J^2 - J(J+1)}{J(2J-1)} \alpha_Q^{(2)}\left(J\right)  \\
     &&  +3\frac{\left(5M_J^2 -J^2 -2J \right)\left(5M_J^2 +1 -J^2 \right) -10M_J^2 \left( 4M_J^2 -1 \right)}
                {J(J-1)(2J-1)(2J-3)}\alpha_Q^{(4)}\left(J\right)
 \label{EQ:ALPHA_Q_SCATEN_GEN}
\end{eqnarray}
In the case of $J=\frac{1}{2}$ and $M_J=\pm \frac{1}{2}$ we have $\alpha_Q\left(\psi_0(M_J)\right) = \alpha_Q^{(0)}\left(J\right)$
and only $\alpha_Q\left(\psi_0(\frac{1}{2})\right)$ needs to be calculated.

For the Tm atom the three tensor quadrupole polarizabilities can be obtained following Ref. \cite{Fleig_PRA_2025}
by calculating $\alpha_Q\left(\psi_0(M_J)\right)$ for the $^2F_{7/2}$ ground term and $M_J \in \{ 7/2,5/2,3/2 \}$. 
The matrix equation resulting from the expansion in Eq. (\ref{EQ:ALPHA_Q_SCATEN_GEN}) then is
\begin{equation}
 \left( \begin{array}{c} \alpha_Q\left(\psi_0(\frac{7}{2})\right) \\ \alpha_Q\left(\psi_0(\frac{5}{2})\right) \\ 
        \alpha_Q\left(\psi_0(\frac{3}{2})\right)
        \end{array}
 \right) =
 \left( \begin{array}{ccc} 1 & 1 & -54 \\ 1 & \frac{1}{7} & -18 \\ 1 & -\frac{3}{7} & -\frac{18}{7} \end{array}
 \right)
 \left( \begin{array}{c} \alpha_Q^{(0)}\left(\frac{7}{2}\right) \\ \alpha_Q^{(2)}\left(\frac{7}{2}\right)  \\ 
        \alpha_Q^{(4)}\left(\frac{7}{2}\right) 
        \end{array}
 \right)
\end{equation}
In terms of the exact inverse matrix this equation becomes
\begin{equation}
 \left( \begin{array}{ccc} -\frac{11}{10} & \frac{7}{2} & -\frac{7}{5} \\ 
                           -\frac{21}{10} & 7 & -\frac{49}{10} \\ 
                           -\frac{7}{90} & \frac{7}{36} & -\frac{7}{60} \end{array}
 \right)
 \left( \begin{array}{c} \alpha_Q\left(\psi_0(\frac{7}{2})\right) \\ \alpha_Q\left(\psi_0(\frac{5}{2})\right) \\ 
        \alpha_Q\left(\psi_0(\frac{3}{2})\right)
        \end{array}
 \right) =
 \left( \begin{array}{c} \alpha_Q^{(0)}\left(\frac{7}{2}\right) \\ \alpha_Q^{(2)}\left(\frac{7}{2}\right)  \\ 
        \alpha_Q^{(4)}\left(\frac{7}{2}\right) 
        \end{array}
 \right)
 \label{EQ:ALPHA_Q_SCATEN_TM_ALL}
\end{equation}
Thus, the scalar quadrupole polarizability for the Tm electronic ground state is expressed as
\begin{eqnarray}
 \alpha_Q^{(0)}\left(7/2\right) &=& -\frac{11}{10} \alpha_Q\left(\psi_0(7/2)\right) + \frac{7}{2} \alpha_Q\left(\psi_0(5/2)\right)  
                                                              -\frac{7}{5}  \alpha_Q\left(\psi_0(3/2)\right)
 \label{EQ:ALPHA_Q_SCATEN_TM7}
\end{eqnarray}
 For the first excited term $^2F_{5/2}$ the ensemble $M_J \in \{ 5/2,3/2,1/2 \}$ is used. 
The matrix equation resulting from the expansion in Eq. (\ref{EQ:ALPHA_Q_SCATEN_GEN}) in this case is
\begin{equation}
 \left( \begin{array}{c} \alpha_Q\left(\psi_0(\frac{5}{2})\right) \\ \alpha_Q\left(\psi_0(\frac{3}{2})\right) \\ 
        \alpha_Q\left(\psi_0(\frac{1}{2})\right)
        \end{array}
 \right) =
 \left( \begin{array}{ccc} 1 & 1 & -98 \\ 1 & -\frac{1}{5} & -18 \\ 1 & -\frac{4}{5} & 4 \end{array}
 \right)
 \left( \begin{array}{c} \alpha_Q^{(0)}\left(\frac{5}{2}\right) \\ \alpha_Q^{(2)}\left(\frac{5}{2}\right)  \\ 
        \alpha_Q^{(4)}\left(\frac{5}{2}\right) 
        \end{array}
 \right)
\end{equation}
In terms of the exact inverse matrix this equation becomes
\begin{equation}
 \left( \begin{array}{ccc} -\frac{19}{27} & \frac{31}{9} & -\frac{47}{27} \\ 
                           -\frac{55}{54} & \frac{85}{18} & -\frac{100}{27} \\ 
                           -\frac{1}{36} & \frac{1}{12} & -\frac{1}{18} \end{array}
 \right)
 \left( \begin{array}{c} \alpha_Q\left(\psi_0(\frac{5}{2})\right) \\ \alpha_Q\left(\psi_0(\frac{3}{2})\right) \\ 
        \alpha_Q\left(\psi_0(\frac{1}{2})\right)
        \end{array}
 \right) =
 \left( \begin{array}{c} \alpha_Q^{(0)}\left(\frac{5}{2}\right) \\ \alpha_Q^{(2)}\left(\frac{5}{2}\right)  \\ 
        \alpha_Q^{(4)}\left(\frac{5}{2}\right) 
        \end{array}
 \right)
 \label{EQ:ALPHA_Q_SCATEN_TM_ALL5}
\end{equation}
The scalar quadrupole polarizability for the Tm first electronic excited state is expressed as
\begin{eqnarray}
 \alpha_Q^{(0)}\left(5/2\right) &=& -\frac{19}{27} \alpha_Q\left(\psi_0(5/2)\right) + \frac{31}{9} \alpha_Q\left(\psi_0(3/2)\right)  
                                                              -\frac{47}{27}  \alpha_Q\left(\psi_0(1/2)\right)
 \label{EQ:ALPHA_Q_SCATEN_TM5}
\end{eqnarray}
The rank-$n$ tensor quadrupole polarizabilities are in the present work obtained by calculating the required quadrupole 
polarizabilities for the projection quantum numbers, i.e., in linear-symmetry representation and then using the above
matrix equations for transformation.

\section{Applications and results}
\label{SEC:RESULTS}

All present calculations have been carried out with a locally-modified version of the DIRAC program package
\cite{DIRAC_JCP}. The matrix elements appearing in Eq. (\ref{EQ:ALPHAQ}) have been implemented in the module
KRCI \cite{knecht_luciparII,fleig_gasci2} that is also used for determining the required atomic wavefunctions.

\subsection{Lithium}

The lithium atomic ground state $^2S_{1/2} (2s^1)$ has been studied earlier and benchmark results have been obtained
that serve for comparison with the results from the present approach.

\subsubsection{Technical Details}

An uncontracted aug-cc-pV5Z+diffuse basis set from the EMSL library \cite{EMSL-basis2019} comprising 
\{$15s,8p,5d,4f,3g,2h$\} functions is used. The Dirac spinors are optimized by diagonalizing a Fock operator with a
fractional occupation of $f = \frac{1}{2}$ per $2s$ spinor. Based on this Dirac-Coulomb-Hartree-Fock (DCHF) 
atomic spinor basis the employed full 
CI model FCI3 comprises all single, double and triple (SDT) excitations of the three electrons into the virtual shells 
below a cutoff of $100$ \au

For the basis set 5Z+ the $d$ space ($\ell = 2$) of the above-described basis set has been singly densified according 
to the basis set extension protocol described in Ref. \cite{Hubert_Fleig_2022}. The resulting $d$ space then comprises
$11d$ functions.

\subsubsection{Results}

\begin{table}[h]
        \caption{\label{TAB:ELQPOL_LI}
           Static electric quadrupole polarizabilities (in \au) for Li ${^2S}_{1/2} (2s^1)$; 
           $K << dim\left(\Gamma(1/2)\right) = 178,843$ (for 5Z basis set)
           is the number of lowest-energy CI eigenvectors forming a subset of the states spanning $\Gamma(1/2)$.
        }

 \vspace*{0.3cm}
 \hspace*{-1.3cm}
 \begin{tabular}{l|l|l}
                                        & \multicolumn{2}{c}{$\alpha_Q$}    \\
    CI model/K                          &       5Z           &      5Z+            \\ \hline
    FCI3\_100au/10                      &     $1424.65$      &    $1425.1$         \\
    FCI3\_100au/20                      &     $1424.74$      &    $1427.3$         \\
    FCI3\_100au/24                      &                    &    $1427.3$         \\
    FCI3\_100au/28                      &                    &    $1427.4$         \\
    FCI3\_100au/30                      &     $1424.74$      &                     \\ \hline
    Lit. \cite{porsev_derevianko_JCP2003} &   \multicolumn{2}{c}{$1424(4)$}        \\
    Lit. \cite{Sternheimer_1970}        &   \multicolumn{2}{c}{$1444.56$}       
  \end{tabular}
\end{table}

Results for lithium are shown in Table \ref{TAB:ELQPOL_LI}. Generally, $\alpha_Q$ converges very rapidly with $K$, the
dimension of the sum over states. As detailed analyses of the excited wavefunctions show the vast majority of the
contribution to $\alpha_Q$ is due to configurational excitations of the type $s \rightarrow d$ which are largely
covered by a small number of contributing excited states. An augmentation of the atomic basis set in the $d$ space
(5Z+ basis) leads to only minor changes in the property results, indicating that the original large basis set denoted
5Z is already sufficiently saturated in the relevant subspaces. 

Finally, the present best results can be compared with literature results. The early more approximate calculation by
Sternheimer \cite{Sternheimer_1970} yields a result that is a few percent removed from the more recent high-level 
calculation by Porsev and Derevianko \cite{porsev_derevianko_JCP2003}. The present results are in excellent 
agreement with the latter calculations. This serves as proof of principle for the present method of calculating
static electric quadrupole polarizability.

\subsection{Thulium}

\subsubsection{Technical Details}

The Dyall QZ basis \cite{dyall_4f} set is used, comprising \{$35s,30p,19d,16f,6g,4h,2i$\} uncontracted functions where 
all $4f,6s,5d$-correlating and $4f$-dipole-polarizing exponents have been added. The Dirac spinors are optimized by 
diagonalizing a Fock operator with a fractional occupation of $f = \frac{13}{14}$ per spinor with $\ell = 3$ 
($4f^{13} 6s^2$ ground configuration). Based on this DCHF atomic spinor basis various CI models are used that are 
defined where they appear. A densified basis set is not used in the case of thulium since E2 polarizability has
proven rather insensitive to such augmentation in the test cases lithium (above) and also in the sodium atom.


\subsubsection{Results}

\begin{table}[h]
        \caption{\label{TAB:ELQPOL_TM7}
           $M_J$-subspace static electric quadrupole polarizabilities (in \au) for Tm ${^2F}_{7/2} (6s^2 4f^{13})$; 
           $K << N = dim\left(\Gamma(M_J)\right)$.
        }

 \vspace*{0.3cm}
 \hspace*{-1.3cm}
 \begin{tabular}{lrcc|c}
    CI model & $K$ & $N$ &  $M_J$parity                      &  $\alpha_Q (M_J)$           \\ \hline
    7s5d\_SDT15 & $20$ & $12,789$ &  7/2u                    &     $3319.5$          \\
                & $30$ &          &                          &     $3321.1$          \\
                & $40$ & &                          &     $3322.5$          \\
                & $50$ & &                          &     $3326.5$          \\
                & $60$ & &                          &     $3326.4$          \\ \hline
    7s5d\_SDT15 & $60$ & $15,232$ &  5/2u                    &     $3378.5$          \\
                & $70$ & &                          &     $3380.3$          \\
                & $120$ &  &                         &     $3380.3$          \\
                & $140$ &  &                         &     $3380.3$          \\ \hline
    7s5d\_SDT15 & $60$ & $17,087$ &  3/2u                    &     $3411.3$          \\
                & $80$ &  &                          &     $3416.4$          \\
                & $100$ &          &                         &     $3416.6$          \\ \hline
    7s5d\_SDT15 & $40$ & $18,093$ &  1/2u                    &     $3429.5$          \\
                & $50$ &          &                          &     $3429.4$          \\
                & $60$ &          &                          &     $3429.5$          \\
                & $70$ &          &                          &     $3429.6$          \\ \hline\hline
    7s5d8s6d\_SDT15 & $150$ & $108,143$ &  7/2u              &     $3243.6$          \\ \hline\hline
    7s5d8s6d9s7d\_SDT15 & $120$ & $371,688$ &  7/2u          &     $3220.6$          \\ \hline\hline
    7s5d8s6d9s7d10s8d\_SDT15 & $100$ & $889,049$ &  7/2u     &     $3194.9$          \\
  \end{tabular}
\end{table}

In a first step a model is used with which reference values for the three tensor polarizabilities can be established,
see Table \ref{TAB:ELQPOL_TM7}.
This model, 7s5d\_SDT15, includes the $15$ electrons from the $6s$ and $4f$ shells and allows for up to triple excitations 
into a virtual space limited to the $7s$ and $5d$ spinors. It thus delivers only a marginal correlation contribution and
is largely equivalent to a Dirac-Hartree-Fock model. The results in Table \ref{TAB:ELQPOL_TM7} demonstrate that with a 
limited number of excited states in the sum over states converged $M_J$-subspace polarizabilites can be obtained.

The models 7s5d8s6d\_SDT15 etc. add further virtual $s$ and $d$ shells to the base model 7s5d\_SDT15. The resulting 
additional excited states formally make positive contributions to $\alpha_Q$ but instead, a reduction of the property 
value is observed. This can be explained by analysis of the individual excited-state contributions. 
The physically exact lowest excited $d$ shell is only approximately
described within the base model 7s5d\_SDT15. Adding the virtual $6d$ spinors leads to an improved description of
the $d$ shell in the excited-state wavefunction, but at the same time to a reduction in $\alpha_Q$ since a $6d$
occupation makes a much smaller contribution to $\alpha_Q$ than a $5d$ occupation and the latter has been reduced.

The ameliorating effect from these extended models is relatively small and makes the calculations much more expensive. 
Therefore, 7s5d\_SDT15 will be used as a base model for calculating correlation effects on $\alpha_Q$ in models that 
to a large degree include the above-described additional excitations (but limited to singles and doubles).

\begin{table}[h]
        \caption{\label{TAB:ELQPOL_TM_TENS}
         Tensor static electric quadrupole polarizabilities (in \au) for Tm ${^2F}_{7/2} (6s^2 4f^{13})$; QZ basis set;
         $^*$ result obtained from correlation-correction model described in the text; $N$ is given for the largest
         irreducible representation $M_J = 3/2$ used in a respective model.
        }

 \vspace*{0.3cm}
 \hspace*{-1.3cm}
 \begin{tabular}{l|r|ccc}
  CI model                  & $dim\left(\Gamma(3/2)\right)$ & $\alpha^{(0)}_Q$  & $\alpha^{(2)}_Q$  & $\alpha^{(4)}_Q$  \\ \hline
  7s5d\_SDT15               &         $17,087$              &     $3389$           &   $-64.7$             &   $-0.0433$       \\
  7s5d\_SDT15\_SD15/1au$^*$ &      $5,887,026$              &     $2727$           &                       &                   \\
  7s5d\_SDT15\_SD15/2au$^*$ &     $15,092,065$              &     $2496$           &                       &                   \\
 \end{tabular}
\end{table}

Using equations (\ref{EQ:ALPHA_Q_SCATEN_TM_ALL}) and (\ref{EQ:ALPHA_Q_SCATEN_TM7}) the tensor quadrupole polarizabilities 
$\alpha^{(0)}_Q$, $\alpha^{(2)}_Q$, and $\alpha^{(4)}_Q$ can be obtained, shown in Table \ref{TAB:ELQPOL_TM_TENS}.
The first line displays the quasi-DCHF results, obtained from complete calculations within the model 7s5d\_SDT15.

Since sum-over-states calculations are very costly computationally a specific strategy is here adopted for obtaining
correlation corrections to the quasi-DCHF values: There is a coverage of about 99\% (model 7s5d\_SDT15\_SD15/1au) 
of the total property value
from only a few excited-state wavefunctions. In the respective $M_J$ subspaces there are three states for 
$M_J=\frac{7}{2}$, three states for $M_J=\frac{5}{2}$ and five states for $M_J=\frac{3}{2}$, respectively,
that make the bulk contribution.
The determinants mainly composing these excited states can be identified within the different models (correlated and 
quasi-DCHF) which allows for obtaining a differential contribution 
\begin{equation}
 \Delta \alpha_Q^{\text{CI model}}(M_J)_j = \alpha^{\text{CI model}}_Q(M_J)_j - \alpha^{\text{quasi-DCHF}}_Q(M_J)_j
 \label{EQ:DELTA_ALPHA_J}
\end{equation}
from the respective states.
It can be unambiguously calculated by comparing the individual contributions to $\alpha_Q(M_J)$ for root $j$. The
total correlation correction $\Delta\alpha_Q(M_J)$ then results from
\begin{equation}
 \Delta\alpha^{\text{CI model}}_Q(M_J) = \sum\limits_j\, \Delta \alpha^{\text{CI model}}_Q(M_J)_j
\end{equation}
for a given CI model. The final result $\alpha^{\text{CI model}}_Q(M_J)$ is obtained as
\begin{equation}
 \alpha^{\text{CI model}}_Q(M_J) = \alpha^{\text{quasi-DCHF}}_Q(M_J) + \Delta\alpha^{\text{CI model}}_Q(M_J)
\end{equation}
Then, the tensor polarizabilities are calculated from the set of results for $\alpha^{\text{CI model}}_Q(M_J)$ with
$M_J \in \{\frac{7}{2},\frac{5}{2},\frac{3}{2}\}$ and using equation (\ref{EQ:ALPHA_Q_SCATEN_TM_ALL}). 
$\alpha^{(2)}_Q$ and $\alpha^{(4)}_Q$ resulting from the correlation-correction model are not given due to
fluctuations on the order of their total values. The results from the base model, however, can serve as
quasi-DCHF estimates of $\alpha^{(2)}_Q$ and $\alpha^{(4)}_Q$ for Tm ${^2F}_{7/2} (6s^2 4f^{13})$.

CI model = 7s5d\_SDT15\_SD15/1au defines that all single and double excitations from the configurations generated by the
basic model 7s5d\_SDT15 into virtual spinors with energies below 1 \au are added to the CI space defined by 7s5d\_SDT15.
Thus, this corresponds to a limited treatment of electron correlation effects on $\alpha_Q$. In detail, it is observed
that the property contributions of all involved majorly contributing excited states are diminished, leading to a total
reduction of $\alpha^{(0)}_Q$ by almost 20\%. Physically, this reduction can be rationalized as follows. The uncorrelated
model 7s5d\_SDT15 overestimates the occupation of the excited $d$ shells in the relevant excited states relative to the
exact wavefunctions for those states, leading to an overestimation of the property value. Including electron correlation
effects corrects for this overestimation of the $d$ occupation by shifting electron density to the orthogonal complement,
i.e., energetically higher-lying virtual states. This shift of electron density is responsible for the reduction of the 
quadrupole polarizability. 

Increasing the energy cutoff in the virtual spinor space to $2$ \au entails a further substantial reduction of 
$\alpha_Q^{(0)}$ by around $7$ percentage points (relative to the base-model result). However, this correction is
much smaller than the one obtained through the previous model 7s5d\_SDT15\_SD15/1au. Due to increasing energy
denominators a further increase of the energy cutoff in the virtual-complement space is expected to yield further,
but small, downward corrections to $\alpha_Q^{(0)}$. They are not calculated explicitly in the present work due
to sharply increasing computational demand. 

Finally, an uncertainty estimate will be reasoned for the static scalar electric quadrupole polarizability in ground-state
atomic thulium. The present calculations show that the majority of uncertainty is due to an incomplete electron
correlation treatment. Improvements in the employed Hamiltonian such as the Breit term and the accounting for radiative
corrections are expected to be not greater than $\approx 1-2$\% \cite{Fleig_PRA_2025}. Basis-set effects combined
with a more complete account for electron correlation effects may yield corrections of a few percent, but likely
following the trend of making $\alpha_Q^{(0)}$ smaller for the reasons discussed in the present work. Therefore,
the final value assigned to the static scalar electric quadrupole polarizability from the present work is
$\alpha_Q^{(0)}(\frac{7}{2}) = 2496^{+0}_{-231}$ \au, where the maximum downward correction is estimated from the
difference between the present best and second-best computational models.

\begin{table}[h]
        \caption{\label{TAB:ELQPOL_TM5}
           $M_J$-subspace static electric quadrupole polarizabilities (in \au) for Tm ${^2F}_{5/2} (6s^2 4f^{13})$; 
           $K << N = dim\left(\Gamma(M_J)\right)$.
        }

 \vspace*{0.3cm}
 \hspace*{-1.3cm}
 \begin{tabular}{lrcc|c}
    CI model & $K$ & $N$ &  $M_J$parity              &  $\alpha_Q (M_J)$      \\ \hline
    7s5d\_SDT15 &  $70$ & $15,232$ &  5/2u           &     $3325.5$           \\ \hline
    7s5d\_SDT15 &  $80$ & $17,087$ &  3/2u           &     $3387.0$           \\ \hline
    7s5d\_SDT15 &  $70$ & $18,093$ &  1/2u           &     $3418.1$           \\ \hline\hline
  \end{tabular}
\end{table}

Data for $\alpha_Q (M_J)$ in the first excited thulium term ${^2F}_{5/2} (6s^2 4f^{13})$ are displayed in Table
\ref{TAB:ELQPOL_TM5} for the base model only. The calculations are converged with respect to the number of states
in the summation as for the electronic ground term.

\begin{table}[h]
        \caption{\label{TAB:ELQPOL_TM_TENS5}
         Tensor static electric quadrupole polarizabilities (in \au) for Tm ${^2F}_{5/2} (6s^2 4f^{13})$; QZ basis set;
         $^*$ result obtained from correlation-correction model described in the text.
        }

 \vspace*{0.3cm}
 \hspace*{-1.3cm}
 \begin{tabular}{l|ccc}
  CI model                  &  $\alpha^{(0)}_Q$  & $\alpha^{(2)}_Q$  & $\alpha^{(4)}_Q$  \\ \hline
  7s5d\_SDT15               &      $3376$           &   $-52.5$             &   $-0.0194$       \\
 \end{tabular}
\end{table}

Table \ref{TAB:ELQPOL_TM_TENS5} shows the resulting tensor quadrupole polarizabilities for the excited term, obtained
by using Eq. (\ref{EQ:ALPHA_Q_SCATEN_TM5}).
At the quasi-DCHF level (model 7s5d\_SDT15) $\left|\alpha^{(n)}_Q(5/2)\right| < \left|\alpha^{(n)}_Q(7/2)\right|
~~ \forall n$.
However, this difference is extremely small, amounting to only $0.4$\% for $\alpha^{(0)}_Q$. Electron correlation
effects on electric polarizability are unsurprisingly very similar for Tm ${^2F}_{5/2}$ and ${^2F}_{7/2}$ states
\cite{Fleig_PRA_2025}. Therefore, a correlated result for $\alpha^{(0)}_Q(\frac{5}{2})$ is expected to be very close
to the final result obtained for $\alpha^{(0)}_Q(\frac{7}{2})$. A direct calculation of the differential static
quadrupole polarizability is not attempted here, since the present sum-over-states approach has uncertainty much
too great to resolve the difference reliably. This was not the case for the finite-field method employed in Ref.
\cite{Fleig_PRA_2025} which allows for significantly more accurate calculation of the static dipole polarizability.

%
%
%
\section{Conclusions}
\label{SEC:CONCL}
A method for calculating static electric quadrupole polarizability in atomic states of arbitrary electronic structure
is presented. It is based on relativistic configuration interaction wavefunctions in linear symmetry and a 
sum-over-states approach with which highly accurate results can be obtained for test systems, here the lithium
atom.

The method is then applied to the ground and first-excited terms of the thulium atom where accurate calculations
are much more challenging. Electron correlation effects on electric quadrupole polarizability are here obtained
from a differential correction method for the ground electronic term and an estimate is made for the electronically
excited term.

The static electric quadrupole polarizability in the $^2F_{7/2}$ and $^2F_{5/2}$ terms of the thulium atom are 
found to be roughly a factor of four smaller than same property in the atomic ground state of the cesium atom 
\cite{Safronova_quadrupole_2016,porsev_derevianko_JCP2003}. This makes thulium significantly less sensitive to stray
electric-field gradients than cesium, at least as far as the contribution from E2 polarizability is concerned.
The present findings add to the known favorable properties of thulium atoms in terms of their extremely small
electric quadrupole moment \cite{Fleig_PhysRevA.107.032816,fritzsche_Tm_2024} and blackbody shift
\cite{Fleig_PRA_2025} and provide further motivation for the development of an ultra-precise thulium atomic clock.
\clearpage
\newcommand{\Aa}[0]{Aa}

\clearpage

\end{document}